\magnification=1200
\baselineskip=16pt
\vskip 0cm
\centerline{\bf Surface Magnetisation and Surface Correlations}
\bigskip
\centerline{\bf in Aperiodic Ising Models}
\vskip 1cm
\centerline{Ferenc Igl\'oi}
\bigskip
\centerline{Research Institute for Solid State Physics, 
H-1525 Budapest, P.O.Box 49, Hungary}
\smallskip
\centerline {and}
\centerline{Institute for Theoretical Physics, Szeged University}
\centerline{H-6720 Szeged, Aradi V. tere 1, Hungary}
\bigskip
\centerline{and}
\bigskip
\centerline{P\'eter Lajk\'o}
\bigskip
\centerline{Institute for Theoretical Physics, Szeged University}
\centerline{H-6720 Szeged, Aradi V. tere 1, Hungary}
\vskip 1cm
{\bf Abstract:}
\vskip 1cm
We consider the surface critical behaviour of diagonally layered Ising models
on the square lattice
where the inter-layer couplings follow some aperiodic sequence.
The surface magnetisation is analytically evaluated from a simple
formula derived by the diagonal transfer matrix method, while
the surface spin-spin correlations are obtained numerically by a recursion
method, based on the star-triangle transformation. The surface critical
behaviour of different aperiodic Ising models are found in accordance
with the corresponding relevance-irrelevance criterion. For marginal sequences
the critical exponents are continuously varying with the strength of
aperiodicity and generally the systems follow anisotropic scaling at the
critical point.
\vskip 0cm
PACS-numbers: 05.50.+q, 64.60.Ak, 68.35.Rh
\vfill
\eject
{\bf I. Introduction}
\vskip 1cm
The discovery of quasi-crystals[1] has stimulated intensive research to understand
their structure and physical properties (for recent reviews see[2-6]).
Theoretically a challenging problem is
to determine the critical properties of such quasiperiodic or more generally
aperiodic structures. After a series of numerical[7-13] and analytical[14-21]
studies on
specific models Luck has proposed a relevance-irrelevance criterion[22]. 
According to this criterion, which is a generalisation of the Harris criterion
for random magnets[23], the inhomogeneity is irrelevant (relevant) if the
fluctuating energy in the scale of the bulk correlation length is smaller
(greater) than the excess thermal energy. In layered systems the above
criterion is connected to the sign of the cross-over exponent:
$$\phi=1+\nu (\omega -1)~~~, \eqno(1)$$
which is expressed in terms of the correlation length exponent $\nu$
of the unperturbed system and the wandering exponent of the sequence $\omega$[24].

Most of the studies about the critical properties of aperiodic systems are
restricted to the quantum Ising chain with aperiodic couplings[22][25-34],
which is equivalent to the two-dimensional classical, layered Ising
model in the extreme anisotropic limit[35]. According to eq(1) for this model
with $\nu=1$
sequences with bounded (unbounded) fluctuations represent irrelevant
(relevant) perturbations. The obtained analytical and numerical results
on different physical quantities (specific heat, surface and bulk
magnetisation, local energy density, etc)
of different aperiodic Ising quantum chains are consistent with the
prediction of the Luck criterion. For marginal sequences non-universal
critical behaviour was found[30-31], even if the aperiodic perturbation is of
radial symmetry[34]. We note also on some related studies on hierarchical
Ising models[36].

In the present paper we study the surface magnetisation and the surface
spin-spin correlations of aperiodic Ising models. In contrast to previous
investigations here we consider the classical version of the model with a
layered aperiodicity in the diagonal direction and the critical properties
are studied on the (1,1) surface of a square lattice. We use two methods
of investigations. The surface magnetisation is calculated analytically
with the diagonal transfer matrix method[37-39], whereas both the surface
magnetisation and
surface correlations are numerically studied by a recursion method based
on the star-triangle transformation. This latter method has also been
used to study layered triangular systems. We note that some preliminary
results of our investigations have already been announced in a Letter[30].

The structure of the paper is the following. In Sections.2 and 3 we present the
methods of diagonal transfer matrix and that of the star-triangle recursion,
respectively. Results on different aperiodic Ising models are given
in Sec 4. A discussion is contained in Sec 5, while detailes of the
calculation are presented in the Appendix.
\vskip 1cm
{\bf II. Surface Magnetisation by the Transfer Matrix Method}
\bigskip
Let us consider an Ising model on the square lattice with a diagonally
layered structure, where the nearest neighbour couplings in the $i$-th
layer from the surface are given by $J_i=K_i k_B T$ (see Fig 1a). We are
interested in the magnetisation at the (1,1) surface, which is
calculated in the transfer matrix formalism[37-39]. Denoting by
$<0|$ and $<1|$ the ground state and the first excited state of the ${\bf T}$
diagonal transfer matrix, respectively, the surface magnetisation
is given
by the matrix-element of the surface spin-flip operator $\sigma_1^x$, as:
$$m_s=<0|\sigma_1^x|1>~~~.\eqno(1)$$
Working with free boundary conditions ${\bf T}$ is different in odd and
even sites, therefore we consider ${\bf T}^2$, which is given for the
inhomogeneous model as:
$$\left( {\bf T}^2 \right)_{\mu,{\overline{\mu}}}=2^N \prod_{i=1}^{N-1}
\cosh\left[ K_{2i-1}(\mu_i+{\overline{\mu}}_i)+K_{2i}(\mu_{i+1}
+{\overline{\mu}}_{i+1})\right] \cosh\left[ K_{2N-1}(\mu_N+{\overline{\mu}}_N)
\right]~~~,\eqno(2)$$
and depends on the configurations of the $\mu_i=\pm 1$ and
${\overline{\mu}}_i \pm 1$ spins, $i=1,2,\dots N$ (Fig. 1b).

To determine the eigenvectors of ${\bf T}^2$ we make use of the fact that
${\bf T}^2$ and the linear operator
$${\bf H}=-\sum_{i=1}^{N-1} \lambda_i \sigma_i^z \sigma_{i+1}^z -
\sum_{i=1}^N h_i \sigma_i^x \eqno(3)$$
commute
$$\left[ {\bf T}^2,{\bf H} \right]=0~~~,\eqno(4a)$$
if the couplings of the inhomogeneous quantum Ising chain in eq(3)
satisfy the relations:
$$h_i{C_{2i-2} \over C_{2i}}=h_{i+1}{C_{2i+1} \over C_{2i-1}} \eqno(4b)$$
and
$$\lambda_i=h_i S_{2i} S_{2i-1} {C_{2i-2} \over C_{2i}}~~~. \eqno(4c)$$
Here we used the abbreviations $\sinh 2K_i \equiv S_i$, $\cosh 2K_i \equiv C_i$
and $C_0=1$. Derivation of eqs(4a-c) is shown in the Appendix.
According to eq(4a) the eigenvectors of ${\bf T}^2$ and
${\bf H}$ are the same, therefore we evaluate the matrix-element in eq(1)
for the inhomogeneous quantum Ising chain.
Using a free fermionic representation of ${\bf H}$[40]
one can show[41] that:
$$m_s=\Phi_s(1)~~~,\eqno(5)$$
and the $\Phi_s$ vector is determined by the equation
$({\bf A}+{\bf B})\Phi_s=0$, where
$${\bf A}+{\bf B}=\pmatrix{h_1&\lambda_1&&&&&\cr
                           &h_2&\lambda_2&&&&\cr  
                           &&h_3&\lambda_3&&&\cr
                           &&&\ldots&&\cr
                           &&&&\ldots&\cr
                           &&&&&\cr} \eqno(6)$$
From the normalization condition $\sum_i \Phi_s^2(i)=1$ one obtains for
the surface magnetisation
$$m_s=\left[1+\sum_{i=1}^{\infty} \prod_{j=1}^i \left( {h_j \over \lambda_j}
\right)^2 \right]^{-1/2}=\left[1+\sum_{i=1}^{\infty} \prod_{j=1}^i S_j
^{-2} \right]^{-1/2}~~~,\eqno(7)$$
where in the last equation we used eqs(4b) and (4c). Analysing the formula
in eq(7) we can say that the magnetisation on the (1,1) surface of a
diagonally layered Ising model is formally the same as that of a quantum
Ising chain with inhomogeneous couplings $\lambda_i=\sinh 2K_i$ and in
uniform transverse field $h_i=1$[41]. We note that for the homogeneous Ising
model eq(7) gives Peschel's result[42]: $m_s=(1-\sinh^{-2} 2K)^{-1/2}$.
\vskip 1cm
{\bf III. Recursion Method}
\bigskip
The Ising model on the triangular lattice is invariant under the star-triangle
transformation[43] (STT), which makes the exact solution of the model on this lattice
relatively simple[44]. Also an exact renormalization group transformation for
the triangular Ising model is based on the repeated use of the STT[45]. 
For a semi-infinite Ising model the STT has been used by Hilhorst and van
Leeuwen[46] and by others[47,48] to construct an iterative procedure to
calculate the
surface magnetisation and the surface correlations in the triangular
Ising model. The method can be succesfully used for layered systems in which
the couplings are the same within one layer. We note that the square lattice
can be considered as a special case of the triangular lattice with vanishing
couplings across the diagonals (Fig. 1a). In the following we shortly
recapitulate the basic results of the recursion method.

Let us consider a layered Ising model on a semi-infinite triangular lattice
with vertical couplings parallel to the surface $\overline{K}_i$,
$i=1/2,3/2,\dots$
and with diagonal couplings $K_i$, $i=1,2,\dots$ (Fig 1a). The STT maps the
triangular lattice onto a hexagonal lattice which is in turn equivalent to a
new triangular lattice. Iterating this mapping a sequence of triangular
Ising models is generated ($n=0,1,2,\dots$) with couplings
${\overline{K}}_i(n)$ and
$K_i(n)$ from the original model with $n=0$. The surface magnetisation
$m_s(n)$ and the surface spin-spin correlation function
$g_s(l,n)=<\sigma_{1,l}\sigma_{1,0}>-<\sigma_1>^2$ transform as[46]:
$$m_s(n)=\left\{ 1-\exp \left[-4\overline{K}_{1/2}(n+1)\right]\right\}^{1/2}
m_s(n+1)~~~,\eqno(8a)$$
$$\eqalign{g_s(l,n)=&{1 \over 4}\left\{ 1-\exp \left[-4\overline{K}_{1/2}(n+1)
\right]\right\}\cr
\times &\left[g_s(l+1,n+1)+2g_s(l,n+1)+g_s(l-1,n+1)\right]~.\cr}\eqno(8b)$$
Making use of the boundary condition $g_s(0,n)=1-m_s^2(n)$ one obtains for
the original model with $m_s=m_s(0)$ and $g_s(l)=g_s(l,0)$[46]:
$$\eqalign{m_s&=\lim_{n\to\infty} [f(n)]^{1/2} m_s(n)~~~,\cr
g(l)&=\sum_{n=1}^{\infty}4^{-n}{l \over n} {2n \choose n+l} f(n)[1-m_s^2(n)]
~~~,\cr
f(n)&=\prod_{j=1}^n\left\{ 1-\exp \left[-4\overline{K}_{1/2}(n+1)\right]
\right\}~~~.\cr}\eqno(9)$$
These relations are exact and can be used to iterate on a computer for any
type of distribution of the couplings in the original layered model. In this
way calculating the surface magnetisation one can numerically determine the
$T_c$ critical point and the $\beta_s$ critical exponent of the surface
magnetisation of the model from $m_s(t) \sim t^{\beta_s}$ as
$t=(T_c-T)/T_c \to 0$. For the square lattice with $\overline{K}_i(0)=0$
these results should be compared with the analytical expression in eq(7).

To obtain analytical results by the recursion method one should analyse the
asymptotic behaviour of $X(n)=\exp[-4\overline{K}_{1/2}(n)]$, since according
to numerical
observations $X(n)$ is smoothly varying with $n \gg 1$[48]. Inserting the
asymptotic solution of $X(n)$ into eqs(9) one obtains in the continuum
approximation:
$$m_s=\left[f(n_0)\right]^{1/2} \exp\left[-{1 \over 2} \int_{n_0}^{\infty}
X(n) dn \right] \eqno(10)$$
and
$$g_s(l)=\sum_{n=l}^{\infty}{l \over n^{3/2}}{1 \over \sqrt{n}}
\exp(-l^2/n) \left[f(n)-f(\infty)\right]~~~,\eqno(11)$$
where $n_0$ is a finite cutoff, on which the critical exponents do not
depend.
According to eq(10) the surface magnetisation is non-zero, if the integral
$\int_{n_0}^{\infty} X(n) dn$ is convergent, thus $X(n)$ goes to zero
faster than $1/n$, as $n$ tends to infinity.

The asymptotic behaviour of $X(n)$ has been calculated exactly at the critical
point of the homogeneous
model and for models with smoothly varying couplings at the surface[46-48],
but there are no exact results available on $X(n)$ outside the critical point.
For the homogeneous, critical Ising model[46]
$X(n)\simeq 1/2n$, thus $m_s(t=0)=0$, $f(n) \sim n^{-1/2}$ and the surface
correlations from eq(11) decay as $g_s(l) \sim l^{-\eta_{\parallel}}$ with
$\eta_{\parallel}=1$. For general inhomogeneous models the decay exponent
follows from the asymptotic behaviour:
$$\lim_{n \to \infty} 2n X(n)=\eta_{\parallel}~~. \eqno(12)$$ 
In numerical calculations it is more accurate to determine the decay
exponent from eq(12), than to investigate the
magnetisation exponent $\beta_s$ from the behaviour of the surface
magnetisation outside the critical point.
\vskip 1cm
{\bf IV Results on Aperiodic Models}
\bigskip
Although one can study general, triangular Ising models by the recursion
method, here we restrict ourselves to the (1,1) surface of diagonally
layered square models.  
In this way we reduce the space of parameters with $\overline{K}_i=0$,
furthermore we make use of the analytical expression
on the surface magnetisation in eq(7).

The criticality condition for layered inhomogeneous Ising models[49] is
expressed in terms of the variable $S_i=\sinh 2K_i$ as:
$$\lim_{L\to \infty} {1 \over L} \sum_{i=1}^L \log S_i=0 \eqno(13)$$
Here we study two-valued sequences of the couplings and use the
parametrization $S_i=S r^{f_i}$,
where $f_i$ takes the values 0 or 1 according to an aperiodic sequence.
The homogeneous model is described by $r=1$. The fluctuation
of the couplings in a domain of size $L$ is characterised by the cumulated
deviation from the average value $\overline{S}$ as[50]:
$$\Delta (S)=\sum_{i=1}^L(S_i-\overline{S}) \approx \delta L^{\omega} F\left(
{\ln L \over \ln \Lambda_1} \right)~~~. \eqno(14)$$
Here $\delta=\overline{S}(r-1)$ is the amplitude of the modulation, $\omega$ is
the wandering exponent, which is expressed by the leading eigenvalues of the
substitutional matrix[24] $\omega=\ln|\Lambda_2|/\ln \Lambda_1$ and $F(x)$ is a
fractal function of its argument with period unity. From the transformation
law of $\delta$ under scaling one can obtain the crossover exponent $\phi$[25]
in eq(1) and the corresponding relevance-irrelevance criterion as described
in the Introduction. The aperiodic sequences we consider in the following
represent different types of perturbation according to this
relevance-irrelevance criterion.
\bigskip
{\it A. Irrelevant perturbation: Thue-Morse sequence}
\bigskip
The binary Thue-Morse sequence[51] is generated through the substitution
$0 \to 01$ and $1 \to 10$, so that one obtains after four steps:
$$0~1~1~0~1~0~0~1~1~0~0~1~0~1~1~0~~.$$
This sequence represents an irrelevant perturbation, since $\Lambda_2=0$ and
$\omega=-\infty$.

The surface magnetisation can be obtained from eq(7) using the corresponding
result for the Thue-Morse quantum Ising chain in Ref[27]:
$$m_s={2 t^{1/2} \over r^{1/2}  + r^{-1/2}}\left[1+{1 \over 4}
\left({r-1 \over r+1}\right)^2 t +O(t^2)\right]~~,\eqno(15)$$
where the critical point is at $S_c=r^{-1/2}$ and $t=1-(S_c/S)^2$.
The surface magnetisation exponent $\beta_s=1/2$ takes the value for
homogeneous Ising systems. Similar conclusion can be obtained from a
study of surface critical correlations. According to numerical results
the relation in eq(12) $lim_{n\to \infty} 2nX(n)=\eta_{\parallel}=1$ is
satisfied with an
accuracy of $10^{-5}$. Thus also the decay exponent takes the value for
homogeneous Ising systems in two dimensions and the perturbation is
indeed irrelevant as expected from scaling.
\bigskip
{\it B. Relevant perturbation: Rudin-Shapiro sequence}
\bigskip
The Rudin-Shapiro sequence[51] is generated by the two digit substitution
$00 \to 0001$, $01 \to 0010$, $10 \to 1101$ and $11 \to 1110$, thus one
obtains after three substitutions:
$$0~0~0~1~0~0~1~0~0~0~0~1~1~1~0~1.$$
The wandering exponent of the sequence is $\omega=1/2$, thus according to
eq(1) this type of perturbation is relevant for the Ising model. The
critical point from eq(13) is given by $S_c=r^{-1/2}$ as for the Thue-Morse
model. The surface magnetisation is again obtained from eq(7) using the
known results about the corresponding Ising quantum chain in Ref[28].
The surface magnetisation behaves differently for $r<1$ and $r>1$. For
$r<1$, when the couplings at the surface are locally stronger than the
average the surface stays ordered at the critical point and the surface phase
transition is of first order. The critical surface magnetisation is given
by[28]:
$$m_{s,c}={1-r \over \sqrt{1-r+r^2}}~~~~~r \le 1~~. \eqno(16)$$
In the other regime, $r>1$, the couplings are locally weaker at the surface
than in the bulk and the surface magnetisation behaves anomalously, it has
an essential singularity at the critical point:
$$m_s \sim \exp[-const (r-1)^2 t^{-1}]~~~~~r>1. \eqno(17)$$
According to numerical results the decay of critical surface correlations is
also
anomalous. The quantity $lim_{n\to \infty} 2n X(n) \to \infty$, thus according
to eq(12)
at the critical point the surface correlations decay faster than any power and
$g_s(l)$ has a stretched exponential dependence on $l$.
\bigskip
{\it C. Marginal perturbations}
\smallskip
{\it Fredholm sequence}
\smallskip
The Fredholm sequence[51] is generated through substitution of the three
letters $A$, $B$ and $C$ as $A \to AB$,
$B \to BC$, $C \to CC$ and we associate $f_i=0$ to the letters $A$ and $C$
and $f_i=1$ to $B$. Starting with a letter $A$ we get for the $f_i$ series
after four substitutions:
$$0~1~1~0~1~0~0~0~1~0~0~0~0~0~0~0~~.$$
This type of perturbation is localised to the surface and there is no
change in the critical temperature, thus $S_c=1$. The sequence is marginal,
since the corresponding wandering exponent $\omega=0$. To evaluate the formula
in eq(7) for the surface magnetisation we use Ref[33]. For $r>\sqrt{2}$ the
surface transition is of first order, the critical surface magnetisation
is given by:
$$m_{s,c}=\sqrt{{r^2-2 \over 2r^2 -3}}~~~,~~~ r\ge \sqrt{2}~~. \eqno(18)$$
In the other regime $r<\sqrt{2}$ the surface transition is of second order and
the corresponding surface magnetisation exponent is a continuous function
of the parameter $r$:
$$\beta_s={1 \over 2} - {\ln r \over \ln 2}~~~,~~~r\le\sqrt{2}~~.\eqno(19)$$
Using the recursion method we have determined the decay exponent of suface
correlations from eq(12), which is shown on Fig 2  for $r \le \sqrt{2}$.
Comparing $\eta_{\parallel}(r)$ with the surface magnetisation exponent
$\beta_s(r)$ in eq(19) we can say that the surface scaling law[52]
$$\eta_{\parallel}=2 \beta_s/\nu \eqno(20)$$
is satisfied for $r \le \sqrt{2}$.
\bigskip
{\it Period-doubling sequence}
\smallskip
The period-doubling sequence follows from the substitution[51] $1 \to 10$ and
$0 \to 11$, so that starting with a $1$ after four steps we have
$$1~0~1~1~1~0~1~0~1~0~1~1~1~0~1~1~.$$
The critical temperature from eq(13) is $S_c=r^{-2/3}$, furthermore the sequence
is marginal since $\omega=0$. The critical exponent of the surface
magnetisation can be analytically
determined using eq(7) and the corresponding result for the Ising quantum chain
in Ref[27]:
$$\beta_s={\ln\left[(1+r^{2/3})(1+r^{-2/3})\right] \over 4 \ln 2 }~~.\eqno(21)$$
As seen from eq(21) $\beta_s(r)$ is continuously varying with the parameter $r$,
furthermore it is the same at both ends of the chain. This is a consequence
of the fact that omitting the last digit the period-doubling sequence is
symmetric.

Next we calculate the decay exponent of critical correlations by the recursion
method. In contrast to the surface magnetisation exponent the decay exponent
is found to be $\eta_{\parallel}=1$, independently of the inhomogeneity
parameter $r$. On Fig. 3  we show for $r=2$ the quantity $2nX(n)$ as a function of
the logarithm of the iterations. Its limiting value as $n \to \infty$ gives
the decay exponent according to eq(12). The log-periodic oscillations for
large $n$
are a consequence of discrete scaling, which can be observed in other
quantities as well (see eq(14)). We note that the same value of the decay exponent is
found on the right boundary of the system.

Comparing the surface magnetisation exponent in eq(21) and the decay exponent
$\eta_{\parallel}=1$ we can say that the surface scaling law in eq(20) does not
satisfy. We shall come back to clear this point in the Discussion.
\bigskip
{\it Paper-folding sequence}
\smallskip
The paper-folding sequence[51] is obtained by recurrent folding of a sheet of
paper, right over left. After unfolding one obtains a series of up- (1) and
down-folds (0). The same sequence can be generated using the two-letter
substitutions $00 \to 1000$, $01 \to 1001$, $10 \to 1100$ and $11 \to 1101$.
Starting with $11$ after three substitutions the sequence is given by:
$$1~1~0~1~1~0~0~1~1~1~0~0~1~0~0~1~.$$
This sequence is also marginal, since $\omega=0$, furthermore the critical
point from eq(13) is $S_c=r^{-1/2}$.

Again the surface magnetisation exponent is analytically known from eq(7)
and using the result for the corresponding quantum Ising chain in Ref[31].
At the left surface
$$\beta_s={\ln(1+r^{-1}) \over 2 \ln 2}~~,\eqno(22)$$
whereas at the right boundary
$$\overline{\beta}_s={\ln(1+r) \over 2 \ln 2}~~,\eqno(23)$$
which is obtained by exchanging perturbed and unperturbed couplings,
i.e. with $r \to r^{-1}$. Thus for the paper-folding sequence, which is
not inversion symmetric, the surface magnetisation exponents are different
at the two boundaries.

Next we turn to calculate the decay exponent on the left boundary by the
recursion method. Now $\eta_{\parallel}$ is found $r$-dependent and for all $r$
it satisfies the relation:
$$\eta_{\parallel}={2 \beta_s \over \beta_s+\overline{\beta}_s}~~, \eqno(24)$$
and a similar equation is true on the right surface with $\beta_s
\leftrightarrow \overline{\beta}_s$. To illustrate the relation in eq(24)
we show on Fig. 4 for $r=2$ the quantity
$nX(n)(\beta_s+\overline{\beta}_s)/\beta_s$, which tends to unity with
log-periodic oscillations, in accordance with eq(12). We can say that the
surface scaling law in eq(20) is again violated, like
to the period-doubling sequence.
\vskip 1cm
{\bf V Discussion}
\bigskip
In this paper we have studied the surface magnetisation and the surface
correlation function of diagonally layered Ising models on the (1,1)
surface. For different aperiodic distribution of the diagonal couplings
we have obtained exact results for the surface magnetisation exponent
by the diagonal transfer matrix method, whereas the decay of surface
correlations were studied numerically by a recursion method based on the
repeated use of the star-triangle transformation. The obtained results are
in accord with the relevance-irrelevance criterion by Luck[22]. For the
relevant Rudin-Shapiro model first-order surface transition and anomalous
decay of critical surface correlations were observed. For marginal
sequences (Fredholm, period-doubling and paper-folding) non-universal
surface critical behaviour was found, the corresponding surface magnetisation
exponents are continuously varying with the inhomogeneity parameter $r$.
 
The above observations remain valid, if the general triangular lattice
Ising model with couplings $K_i$ and $\overline{K}_i$ is concerned. Then
besides the aperiodicity ratio $r$ another parameter $\overline{K}_i/K_i$
enters into the expressions. In this general case the criticality condition
is also known analytically[49]:
$$\lim_{L\to \infty} {1 \over L} \sum_{i=1}^L \log S_i+2\overline{K}_i=0~~,
\eqno(25)$$
while both the surface magnetisation and the surface correlations have to
be calculated numerically by the recursion method. Our results on the
triangular lattice qualitatively agree with that on the diagonal square
lattice, they satisfy the relevance-irrelevance criterion in eq(1) for
all sequences. For marginal sequences continuously varying critical
exponents were found, which depend on two parameters.
Also the corresponding scaling relations are satisfied, eq(20) for the
Fredholm sequence
and eq(24) for the perid-doubling and paper-folding sequences.

Finally, we come to the point to explain the violation of surface scaling
relation in eq(20) for the period-doubling and paper-folding sequences. 
The observed scaling behaviour in eq(24)
is compatible with anisotropic scaling, when the correlation lengths
parallel with $\xi_{\parallel}$ and perpendicular to the
surfaces $\xi_{\perp}$ are diverging with different exponents, so that
$\xi_{\parallel}
\sim \xi_{\perp}^z$, where $z$ is the anisotropy exponent. According to
anisotropic scaling[53] the critical spin-spin correlation function on the
left surface behave as:
$$g_s(l,t)=b^{-2\beta_s/\nu} g_s(l/b^z,b^{1/\nu}t)~~,\eqno(26)$$
when lengths perpendicular to the surface are rescaled by a factor of $b>1$.
At the critical point $t=0$ the decay exponent is given by $\eta_{\parallel}
=2\beta_s/\nu z$, which corresponds to the relation in eq(24), if
$$z=\beta_s+\overline{\beta}_s~~~.\eqno(27)$$
For the period-doubling sequence with $\beta_s=\overline{\beta}_s,~~
\eta_{\parallel}=1$, as observed. We note that the anisotropy exponent $z$
has been recently analytically calculated for the corresponding Ising quantum
chains[54] in accordance with eq(27).
Thus we can conclude that for marginally aperiodic layered Ising models where
the perturbation extends over the volume of the system the systems become
essentially anisotropic at the critical point and the anisotropy exponent can
be expressed as the sum of the two surface magnetisation exponents.
\bigskip
{\bf Acknowledgement}: F.I. thanks for collaborations with L. Turban and M.
Henkel at early stages of this work. He also thanks to I. Peschel for a
useful correspondence.
This work has been supported by the Hungarian National
Research Fund under grant No OTKA TO12830.

\vfill
\eject
{\centerline{\bf Appendix}}
\bigskip
To prove eq(4) we start with the representation of ${\bf H}$ in eq(3) in the
$\mu,{\overline{\mu}}$ basis:
$$H_{\mu,{\overline{\mu}}}=-\sum_{i=1}^{N-1} \lambda_i \mu_i \mu_{i+1}
\delta_{\mu,{\overline{\mu}}}+\sum_{i=1}^N h_i \delta(\mu_i+{\overline{\mu}}_i)
\prod_{j\ne i} \delta(\mu_j-{\overline{\mu}}_j)~~~.\eqno(A1)$$
Then the matrix-elements of the commutator $\left[ {\bf T}^2,{\bf H} \right]$
are given by:
$$\eqalign{ \bigl( {\bf T}^2 {\bf H}& -{\bf H}{\bf T}^2 \bigr)
_{\mu,{\overline{\mu}}}=
-{\bf T}^2_{\mu,{\overline{\mu}}} \Biggl\{ 
\sum_{i=1}^N h_i {\cosh(K_{2i-1}(\mu_i-{\overline{\mu}}_i)
+K_{2i}(\mu_{i+1}
+{\overline{\mu}}_{i+1}))  \over \cosh(K_{2i-1}(\mu_i
+{\overline{\mu}}_i)+K_{2i}(\mu_{i+1}
+{\overline{\mu}}_{i+1}))} \cr
&\times {\cosh(K_{2i-3}(\mu_{i-1}
+{\overline{\mu}}_{i-1})+K_{2i-2}(\mu_{i}
-{\overline{\mu}}_{i}))\over \cosh(K_{2i-3}(\mu_{i-1}
+{\overline{\mu}}_{i-1})+K_{2i-2}(\mu_{i}
+{\overline{\mu}}_{i}))} + \sum_{i=1}^{N-1} \lambda_i
{\overline{\mu}}_i {\overline{\mu}}_{i+1}
-(\mu_i \leftrightarrow {\overline{\mu}}_i) \Biggr\} \cr} \eqno(A2)$$
Here in the surface terms $K_0=K_{-1}=K_{2N}=0$.
The term in the first sum in the r.h.s. of eq(A2) can be rewritten using the identities
 $\sinh[a(\mu\pm
{\overline{\mu}})]=(\mu\pm{\overline{\mu}})/2 \sinh 2a$ and $\tanh[a(\mu\pm
{\overline{\mu}})]=(\mu\pm{\overline{\mu}})/2 \tanh 2a$ as:
$$\bigl[ \mu_{i+1}\mu_i \tanh 2K_{2i} \sinh 2K_{2i-1} \cosh 2K_{2i-2}
+ \mu_{i}\mu_{i-1} \tanh 2K_{2i-3} \sinh 2K_{2i-2} \cosh 2K_{2i-1}$$
$$+\overline{\mu}_{i+1}\mu_i \tanh 2K_{2i} \sinh 2K_{2i-1} 
\cosh 2K_{2i-2}
+ \mu_{i}\overline{\mu}_{i-1} \tanh 2K_{2i-3} \sinh 2K_{2i-2}
 \cosh 2K_{2i-1}$$
$$-(\mu_i \leftrightarrow {\overline{\mu}}_i)\bigr]/2$$
so that we obtain for the commutator:
$$\eqalign{ \bigl[ {\bf T}^2,{\bf H} \bigr]_{\mu,{\overline{\mu}}}=&
-{\bf T}^2_{\mu,{\overline{\mu}}} \Biggl\{ \sum_{i=1}^{N-1}(
\mu_{i+1}\mu_i-{\overline{\mu}}_{i+1} {\overline{\mu}}_{i})
\biggl[{1 \over 2} S_{2i}S_{2i-1}\biggl(h_i {C_{2i-2} \over C_{2i}}
+h_{i+1} {C_{2i+1} \over C_{2i-1}} \biggr) - \lambda_i \biggr] \cr
&+\sum_{i=1}^{N-1}(
{\overline{\mu}}_{i+1}\mu_i-{{\mu}}_{i+1} {\overline{\mu}}_{i})
{1 \over 2} S_{2i}S_{2i-1}\biggl(h_i {C_{2i-2} \over C_{2i}}
-h_{i+1} {C_{2i+1} \over C_{2i-1}} \biggr) \Biggr\} \cr} \eqno(A3)$$
Then the commutator is zero if eqs(4b) and (4c) are satisfied.
\vfill
\eject
{\bf References}
\bigskip
\item{ [1]} D. Schechtman, I. Blech, D. Gratias and J.W. Cahn,
Phys. Rev. Lett. 53, 1951 (1984) 
\bigskip
\item{ [2]} C.L. Henley, Comments Condens. Matter Phys. 13, 59 (1987)
\bigskip
\item{ [3]} T. Janssen, Phys. Rep. 168, 55 (1988)
\bigskip
\item{ [4]} C. Janot, J.M. Dubois and M. de Boissieu, Am. J. Phys. 57, 972 (1989)
\bigskip
\item{ [5]} P. Guyot, P. Kramer and M. de Boissieu, Rep. Prog. Phys. 54, 1373 (1991)
\bigskip
\item{ [6]} {\it Quasicrystals: The State of the Art}, P. Steinhardt and D. DiVincenzo
eds. (World Scientific, Singapore, 1991)
\bigskip
\item{ [7]} C. Godr\'eche, J.M. Luck and H.J. Orland, J. Stat.
Phys. 45, 777 (1986)
\bigskip
\item{ [8]} Y. Okabe and K. Niizeki, J. Phys. Soc. Japan 57, 1536
(1988)
\bigskip
\item{ [9]} E.S. S\/orensen, M.V. Jari\`c and M. Ronchetti, Phys.
Rev. B44, 9271 (1991)
\bigskip
\item{[10]} S. Sakamoto, F. Yonezawa, K. Aoki, S. Nos\`e and M.
Nori, J. Phys. A22, L705 (1989)
\bigskip
\item{[11]} C. Zhang and K. De'Bell, Phys. Rev. B47, 8558 (1993)
\bigskip
\item{[12]} G. Langie and F. Igl\'oi, J. Phys. A25, L487 (1992)
\bigskip
\item{[13]} Y. Okabe and K. Nizeki, J. Phys. A23, L733 (1990)
\bigskip
\item{[14]} B.M. McCoy and T.T. Wu, Phys. Rev. Lett. 21, 549 (1968)
\bigskip
\item{[15]} C.L. Henley and R. Lipowsky, Phys. Rev. Lett. 59, 1679 (1987)
\bigskip
\item{[16]} C.A. Tracy, J. Phys. A21, L603 (1988)
\bigskip
\item{[17]} F. Igl\'oi, J. Phys. A21, L911 (1988)
\bigskip
\item{[18]} M.M. Doria and I.I. Satija, Phys. Rev. Lett. 60, 444
(1988)
\bigskip
\item{[19]} G.V. Benza, Europhys. Lett. 8, 321 (1989)
\bigskip
\item{[20]} M. Henkel and A. Patk\'os, J. Phys. A25, 5223 (1992)
\bigskip
\item{[21]} Z. Lin and R. Tao, J. Phys. A21, L603 (1988)
\bigskip
\item{[22]} J.M. Luck, J. Stat. Phys. 72, 417 (1993)
\bigskip
\item{[23]} A.B. Harris, J. Phys. C7, 1671 (1974)
\bigskip
\item{[24]} M. Queff\'elec, {\it Substitutional Dynamical
Systems-Spectral Analysis}, Lecture Notes in Mathemathics, Vol. 1294,
A. Dold and B. Eckmann eds. (Springer, Berlin, 1987)
\bigskip
\item{[25]} F. Igl\'oi, J. Phys. A26, L703 (1993)
\bigskip
\item{[26]} L. Turban and B. Berche, Z. Phys. B92, 307 (1993)
\bigskip
\item{[27]} L. Turban, F. Igl\'oi and B. Berche, J. Phys. A27, 6349
(1994)
\bigskip
\item{[28]} F. Igl\'oi and L. Turban, Europhys. Lett. 27, 91 (1994)
\bigskip
\item{[29]} L. Turban, P-E. Berche and B. Berche, J. Phys. A27,
6349 (1994)
\bigskip
\item{[30]} B. Berche, P-E. Berche, M. Henkel, F. Igl\'oi, P.
Lajk\'o, S. Morgan and L. Turban, J. Phys. A28, L165 (1995)
\bigskip
\item{[31]} P-E. Berche, B. Berche and L. Turban, J. Phys. I6, 621 (1996)
\bigskip
\item{[32]} U. Grimm and M. Baake, J. Stat. Phys. 74, 1233 (1994)
\bigskip
\item{[33]} D. Karevski, G. Pal\'agyi and L. Turban, J. Phys. A28, 45 (1995)
\bigskip
\item{[34]} D. Karevski, L. Turban and F. Igl\'oi, J. Phys. A28, 3925 (1995)
\bigskip
\item{[35]} J. Kogut, Rev. Mod. Phys. 51, 659 (1979)
\bigskip
\item{[36]} F. Igl\'oi, P. Lajk\'o and F. Szalma, Phys. Rev. B52, 7159 (1995)
\bigskip
\item{[37]} L. Onsager, in {\it Critical Phenomena in Alloys, Magnets and
Superconductors}, R.E. Mills, E. Ascher, R.I. Jaffee eds. (McGraw-Hill,
New York, 1971) 
\bigskip
\item{[38]} M.J. Stephen and L. Mittag, J. Math. Phys. 13, 1944 (1972)
\bigskip
\item{[39]} W.F. Wolff and J. Zittartz, Z. Phys. B44, 109 (1981)
\bigskip
\item{[40]} E.H. Lieb, T.D. Schultz and D.C. Mattis, Ann. Phys. NY.
16, 406 (1961)
\bigskip
\item{[41]} I. Peschel, Phys. Rev. B30, 6783 (1984) 
\bigskip
\item{[42]} I. Peschel, Phys. Lett. 110A, 313 (1985) 
\bigskip
\item{[43]} I. Syozi, in {\it Phase Transitions and Critical
Phenomena}, Vol. 1, C. Domb and M.S. Green eds. (Academic, London,
1972) 
\bigskip
\item{[44]} R.J. Baxter and I.G. Enting, J. Phys. A11, 2463 (1978)
\bigskip
\item{[45]} H.J. Hilhorst, M. Schick and J.M.J. van Leeuwen, Phys. Rev. B19,
2749 (1979)
\bigskip
\item{[46]} H.J. Hilhorst and J.M.J. van Leeuwen, Phys. Rev. Lett.
47, 1188 (1981)
\bigskip
\item{[47]} T.W. Burkhardt and I. Guim, Phys. Rev. B29, 508 (1984)
\bigskip
\item{[48]} T.W. Burkhardt, I. Guim, H.J. Hilhorst and J.M.J. van Leeuwen,
Phys. Rev. B30, 1486 (1984)
\bigskip
\item{[49]} F. Igl\'oi and P. Lajk\'o, Z. Phys. B99, 281 (1996)
\bigskip
\item{[50]} J.M. Dumont, in {\it Number Theory and Physics}, Springer Proc.
Phys., Vol. 47, J.M. Luck, P. Moussa and M. Waldschmidt eds. (Springer, Berlin,
1990) 
\bigskip
\item{[51]} M. Dekking, M. Mend\'es-France and A. van der Poorten, Math.
Intelligencer 4, 130 (1983)
\bigskip
\item{[52]} K. Binder, in {\it Phase Transitions and Critical
Phenomena}, Vol. 8, C. Domb and J.L. Lebowitz eds. (Academic, London,
1983)
\bigskip
\item{[53]} K. Binder and J.S. Wang, J. Stat. Phys. 55, 87 (1989)
\bigskip
\item{[54]} F. Igl\'oi and L. Turban (unpublished)
\vfill
\eject
Figure Captions:
\vskip 1cm
\item{Fig. 1} (a) Diagonally layered square lattice (full line) and the
corresponding triangular lattice with dashed vertical lines. For the square
lattice the vertical couplings are zero $\overline{K}_i=0$. (b) Portion of
the lattice contained in the square of the diagonal transfer matrix.
\vskip 1cm
\item{Fig. 2} Decay exponent of critical surface spin-spin correlations for
the Fredholm Ising model.
\vskip 1cm
\item{Fig. 3} The quantity $2n X(n)$ as a function of the logarithm of the
iterations for the period-doubling Ising model with $r=2$. The decay exponent
$\eta_{\parallel}=1$ in eq(12) is approached through log-periodic oscillations.
\vskip 1cm
\item{Fig. 4}  The quantity $n X(n)(\beta_s+\overline{\beta}_s)/\beta_s$ as a
function of the logarithm of the
iterations for the paper-folding Ising model with $r=2$. The decay exponent
$\eta_{\parallel}=(\beta_s+\overline{\beta}_s)/\beta_s$ in eq(12) is
approached through log-periodic oscillations.
\vfill
\eject
\end
\end